\documentclass[aps,prl,showpacs,twocolumn]{revtex4}
\usepackage{graphicx}

\input{epsf}
\begin{document}

\title{Electromagnetic field induced suppression of transport \\
through $n$-$p$ junctions in graphene}
\author{M. V. Fistul and K. B. Efetov}
\affiliation{Theoretische Physik III, Ruhr-Universit\"at Bochum, D-44801 Bochum, Germany}
\date{\today}

\begin{abstract}
We study quasi-particle transmission through an $n $-$p$ junction
in a graphene irradiated by an electromagnetic field (EF). In the
absence of EF the electronic spectrum of undoped graphene is
gapless, and one may expect the perfect transmission of
quasi-particles flowing perpendicular to the junction. We
demonstrate that the resonant interaction of propagating
quasi-particles with the component of EF parallel to the junction
induces a \textit{non-equilibrium dynamic gap} $(2\Delta_R)$
between electron and hole bands in the quasi-particle spectrum of
graphene. In this case the strongly suppressed quasi-particle
transmission is only possible due to interband tunnelling. The
effect may be used for controlling transport properties of diverse
structures in graphene, like, e.g., $n$-$p$-$n$ transistors,
single electron transistors, quantum dots, etc., by variation of
the intensity $S$ and frequency $\omega$ of the external
radiation.
\end{abstract}

\pacs{73.63.-b,05.60.Gg,81.05.Uw,73.43.Jn} \maketitle


Recent success in fabrication of graphene samples
\cite{NovGeim,Zhang} has resulted in a stream of publications
devoted to study of this interesting material. The unique
properties of graphene originate from peculiarities of the
electron spectrum. The quasi-particle spectrum $\epsilon (p)$
consists of two valleys, and in each valley there are electron and
hole bands crossing each other at some point. Near these points
the electron spectrum is linear
\begin{equation}
\epsilon _{\pm }(p)=\pm v|\mathbf{p}|,  \label{Spectrum}
\end{equation}%
where $\mathbf{\vec{p}}=\{p_{x},p_{y}\}$ is the quasi-particle
momentum, $v$ is the Fermi velocity (only weakly dependent on the
momentum $p$). The quantum dynamics of quasi-particles can
effectively be described by a Dirac- like equation
\cite{Spectrum}. This spectrum of quasi-particles in graphene has
been experimentally verified by observation of specific gate
voltage dependencies of Shubnikov-de Haas oscillations,
conductivity and quantum Hall effect \cite{NovGeim,Zhang}.

Although being different in details, many interesting phenomena in
graphene have their analogues in conventional two dimensional
systems. For example, one can observe the quantum Hall effect with
a specific structure \cite{NovGeim,Zhang} that agrees with
theoretical predictions derived from the Dirac equation
\cite{Graphene-Theor}. Considering effects of disorder one obtains
not just the localization but an interesting
crossover between the antilocalization and localization behavior \cite%
{suzuura,Efetov}.

At the same time, unique effects specific only for graphene are
also possible. One of the most unusual phenomena is the {\it
reflectionless} transmission through a one-dimensional potential
barrier of arbitrary strength predicted in Refs.
\cite{Falko,Katzn} and recently coined as the \textquotedblleft
Klein paradox\textquotedblright\ \cite{Katzn}. The simplest
experimental setup suggested for studying this effect is a
graphene based $n$-$p$ junction that can be made by split-gate
technique \cite{NovGeim,Falko,Katzn} (see schematic in Fig. 1).
The absence of the backscattering of the massless particles
flowing perpendicular to the barrier is related to the chiral
nature of them and to a phenomenon of \textquotedblleft
isospin\textquotedblright\ conservation \cite{Katzn}.
\begin{figure}[tbp]
\includegraphics[width=1.2in,angle=-90]{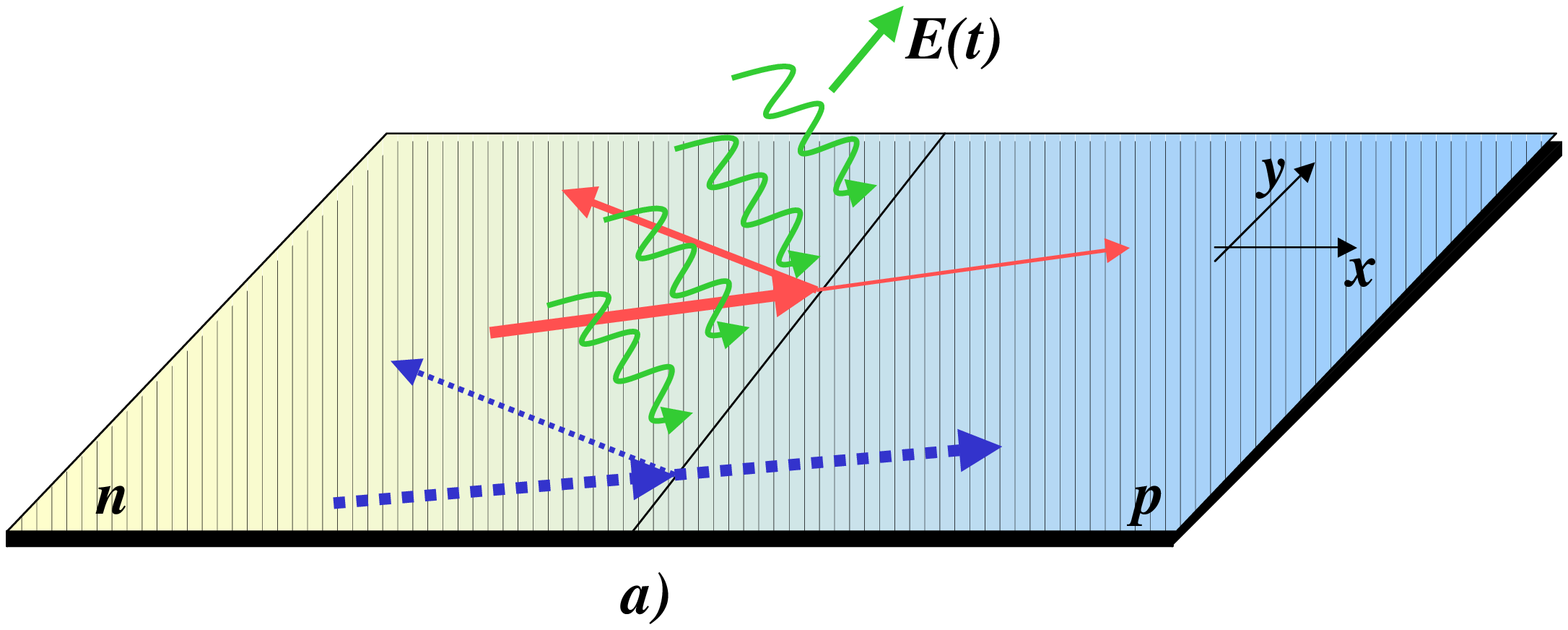}
\caption{An $n$-$p$ junction in graphene. The quasi-particle
scattering in the absence (dashed lines) and in the presence
(solid lines) of the EF is shown. } \label{Schematic}
\end{figure}
The perfect transmission of the quasi-particles can be explained
in a natural way using a standard theory of interband tunnelling
\cite{Tunn}. Indeed, it has been shown that the transmission
probability $P$ through an $n $-$p$ junction is determined by the
gap $\Delta $ between the electron and hole bands as \cite{Tunn}
\begin{equation}
P~\simeq ~\exp [-\frac{\pi \Delta ^{2}}{4\hbar vF}],  \label{a1}
\end{equation}%
where $F$ is the slope of an $x$-dependent electrostatic potential in the $n$%
-$p$ junction. Since the undoped graphene is a gapless material, taking the
limit $\Delta \rightarrow 0$ leads to the conclusion about the ideal
transmission of the quasi-particles flowing perpendicular to the junction.

Such a perfect transmission of quasi-particles through an $n$-$p$ junction
might lead to difficulties in confining electrons in future graphene based
electronic devices (like those, suggested, e.g., in Ref. \cite{Loss}).
Although in narrow stripes this difficulty can be avoided due to transversal
quantization \cite{EfSilv}, the problem may persist in clean wide 2D samples.

At the same time, the reflectionless penetration is rather sensitive to
applying external fields. For example, it is expected \cite{Falko} that a
magnetic field may reduce the transmission or even confine the electrons
\cite{egger} for certain nonhomogeneous configurations of the field. It is
not difficult to imagine that the current flowing through the $n-p$
junctions in graphene may be not less sensitive to other external
perturbations.

\begin{figure}[tbp]
\includegraphics[width=2.6in,angle=0]{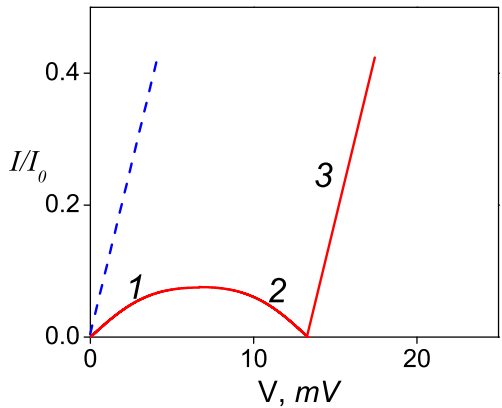} %
\caption{ The CVC of an $n$-$p$ junction: in the presence (solid
line) and the absence (dashed line) of EF. It is calculated by
using Eqs. (\ref{Tunnsupp}) and (\ref {Tunncurrent-2}), and the
parameters $\epsilon_0=0.02 eV$, $d~=1 \mu m$, $\hbar
\omega=(4/3)\epsilon_0$, and $ S=1W/cm^2$.
} \label{CVC}
\end{figure}

In this Letter we predict and analyze a new interesting effect
arising in the $n-p$ junctions formed in a wide graphene,
irradiated by an external electromagnetic field (EF). The system
we consider is represented in Fig. 1.
We show that the radiation leads to a very unusual
dependence of the $dc$ current $I$ on the voltage $V$ applied
across the barrier. At low values of the applied voltage the
dependence is linear but the resistivity extracted from it is much
larger than the one calculated previously \cite{Falko}. At certain
values of the voltage $V$ the current $I$ through the barrier can
even be completely blocked.
The entire EF induced current-voltage characteristics (CVC) having
the $N$-shape is shown in Fig. \ref{CVC}.

We show that the resonant interaction of the Dirac-like particles
with the $y$ component of EF \emph{parallel }to the interface
leads to formation of a \textit{nonequilibrium gap} $(2\Delta
_{R})$ between the electron and hole bands in the quasi-particle
spectrum. This gap and specific non-monotonic dependence of the
quasi-particle transmission on the energy $\epsilon_0$ are
responsible for the unusual form of CVC in Fig. \ref{CVC}.

Formation of such a dynamic gap is well known for two level
systems under radiation \cite{Hanggi}.
Here we have a continuous spectrum rather than just
two levels but the effect persists. Like for the two level
systems, one can have a resonance that can be achieved when the frequency $%
\omega $ of EF satisfies the specific condition $\hbar \omega =2v|\mathbf{p}%
\left( x\right) |$, where $\mathbf{p}\left( x\right) $ is the
coordinate dependent classical momentum of the quasiparticles. The
value of the gap depends strongly on the intensity $S$ and
frequency $\omega $ of external radiation.

The externally applied EF is taken into account in the two bands
time and
coordinate dependent Hamiltonian of the $n$-$p$ junction as follows \cite%
{CoherPhoton}:
\begin{equation}
\hat{H}(t)=v\mathbf{\hat{\sigma} }\left[ \mathbf{\hat{p}-}\frac{e}{c}\mathbf{A}%
\left( t\right) \right] +U(x),  \label{Hamiltonian}
\end{equation}%
where $U(x)$ is the electrostatic potential of the $n$-$p$
junction,
and the $\mathbf{\hat{\sigma}
}=\{\hat{\sigma}_{x},\hat{\sigma}_{y}\}$ are the standard Pauli
matrices in the sublattice space. We assume that the potential
$U(x)$ varies sufficiently slowly, such that the scattering
between different valleys can be neglected. Just for simplicity
and in order to clarify how one comes to the resonant interaction
of quasi-particles with the EF, we consider the case of external
radiation linearly polarized in the
$y$-direction. The electromagnetic wave is characterized by the $y$%
-component of the vector-potential as $A_{y}=(Ec/\omega )\cos
(\omega t)$, and $E=\sqrt{4\pi S/c}$ is the amplitude of the
electric field.

Next we reduce the time-dependent problem described by Hamiltonian (\ref%
{Hamiltonian}) to a stationary problem by switching to the rotating frame
using the following unitary transformation of the two component Dirac wave
functions
\begin{equation}
\hat{U}_n=\frac{1}{\sqrt{2}}\left(
\begin{tabular}{cc}
$1$ & \hspace{0.2cm} $1$ \\
$-\exp (i\hat{\theta})$ & \hspace{0.2cm} $\exp (i\hat{\theta})$%
\end{tabular}%
\ \right) \exp \left[ i\omega t
\left(n-\frac{1}{2}\hat{\sigma}_{z} \right) \right] , \label{a2}
\end{equation}%
where $\hat{\theta}=tan^{-1}(\hat{p}_{y}/\hat{p}_{x})$. The transformed
hamiltonian $\hat{H}_{eff}^{\prime }=\hat{U}^{+}_n H\hat{U}_n-i\hbar \hat{U}^{+}_n%
\hat{\dot{U}}_n$ contains, in general, both static and
proportional to $\exp \left( \pm 2i\omega t\right) $ parts.
However, like for the two level systems \cite{Hanggi}, only the
static part $\hat{H}_{eff}$ is important near
the resonance, and can be written as%
\begin{equation}
\hat{H}_{eff}=\left(
\begin{tabular}{cc}
$\frac{\hbar (2n-1)\omega }{2}+v|\mathbf{\hat{p}}|+U(x)$ & \hspace{0.0cm} $\frac{%
eEv}{2\omega }$ \nonumber\vspace{0.2cm} \\
$\frac{eEv}{2\omega }$ & \hspace{0.0cm} $\frac{\hbar(2n+1) \omega }{2}-v|\mathbf{%
\hat{p}}| +U(x)$%
\end{tabular}%
\ \right) ~,  \label{a3}
\end{equation}%
where $|\mathbf{\hat{p}}|=\sqrt{\hat{p}_{x}^{2}+\hat{p}_{y}^{2}}$.
Neglecting the oscillating part of the Hamiltonian
$\hat{H}_{eff}^{\prime }$ corresponds to a rotation wave
approximation (RWA) \cite{Hanggi}. The RWA is valid in the most
interesting regime of the \textit{resonant interaction} between
the EF and propagating quasi-particles when
\begin{equation}
\hbar \omega \simeq 2v|\mathbf{p}\left( x\right) |  \label{a100}
\end{equation}
A weak non-resonant interaction of quasiparticles with EF
neglected here was studied in Ref. \cite{Nonres}. The most
important contributions come from almost one-dimensional electron
motion, and we assume in our consideration that $p_{x}\gg p_{y}$.
We also assume that the amplitude of the external microwave
radiation is comparatively small, $eEv/\hbar \ll \omega ^{2}$.

The Eq. (\ref{a3}) shows that the radiation results in the
appearance of off-diagonal elements in the operator
$\hat{H}_{eff}$. In the absence of the
coordinate dependent potential, i.e. $U(x)=0$, the eigenvalues $\tilde{%
\epsilon}(p)$ of $\hat{H}_{eff}$ give the sets of bands of
quasi-energies (the Floquet eigenvalues \cite{Hanggi}):
\begin{equation}
\tilde{\epsilon}_{n,\pm}(p)~=~n\omega \pm
\sqrt{(v|\mathbf{p}\left( x\right) |-\frac{%
\hbar \omega }{2})^{2}+\Delta _{R}^{2}} \label{Eigenvalues}
\end{equation}%
where
\begin{equation}
 2\Delta _{R}=(ev/\omega) \sqrt{4\pi S/c} \label{a4}
\end{equation}%
is the EF induced non-equilibrium gap. The $n$ are  an integer
number $0, \pm 1, \pm 2,...$.

It is well known \cite{Hanggi} that, in the presence of periodic
time-dependent perturbations, the bands of the Floquet eigenvalues replace
the quasi-particle spectrum, Eq. (\ref{Spectrum}). The quantity $\Delta
_{R}/\hbar $ has the same meaning as the famous Rabi frequency for microwave
induced quantum coherent oscillations between two energy levels (these
energy levels are $v|\mathbf{p}\left( x\right) |$ and $-v|\mathbf{p}\left(
x\right) |$ in our case).

Next we analyze the quasi-energy $\epsilon _{0}$ dependent
transmission of quasi-particles $P(\epsilon _{0})$ through the
potential barrier $U(x)$ formed in the $n$-$p$ junction. To obtain
the analytical solution we use the quasi-classical approximation
that can be quite realistic for the $n$-$p$ junctions created
electrostatically. The classical phase trajectories $p(x)$ of the
Hamiltonian $\hat{H}_{eff}$ are determined by the conservation of
the sum of the potential energy $U(x)$ and the quasi-energy
$\tilde{\epsilon}_{n,\pm}(p)$ as
\begin{equation}
U(x)+\tilde{\epsilon}_{n,\pm}(p)=\epsilon_0~~.
\label{energyconserv}
\end{equation}%
Using Eq. (\ref{energyconserv}) for $n=0, \pm 1$ we obtain three
regimes of the quasi-particle propagation through the barrier
depending on the energy of electrons $\epsilon _{0}$ and frequency
$\omega$. As $\epsilon _{0}>\hbar \omega /2$, the momentum $p_{x}$
decreases as the quasi-particle approaches the barrier, the
resonant condition is satisfied sufficiently close to the junction
and, therefore, the interaction with the EF results in the
classical reflection of the quasi-particle. This is shown
schematically in Fig. 3a by a thick solid line. The transmission
of the particles through the barrier occurs in the form of the
\textit{non-equilibrium interband tunnelling} between electron and
hole Floquet bands, and therefore, the effect is a particular
example of the dynamical tunnelling \cite{Hanggi}. The
quasiparticles tunnel from the electronic $\tilde{\epsilon}_{e}$-
band ($n=0$) on the left side to the hole $\tilde{\epsilon}_{h}$-
band ($n=0$) on the right side of the junction. Similarly to the
usual case of the interband tunnelling \cite{Tunn} the probability
$P(\epsilon _{0})$ of the quasi-particle transmission is
determined by the following process in the \textquotedblleft under
barrier region\textquotedblright\
: the quasi-particle moves from the left \textquotedblleft
classical turning point\textquotedblright\ ($p=\hbar \omega
/(2v)$) to \textquotedblleft the branch point\textquotedblright\
($p=\hbar \omega /(2v)+i\Delta _{R}/v$) in the complex $(x,p)$
plain, and afterwards to the right \textquotedblleft classical
turning point\textquotedblright.

A further progress can be made by choosing a specific model for
the electrostatic potential of voltage biased $n$-$p$ junction
($d$ is the $n$-$p$ junction width) \cite{Tunn}
\begin{equation}
U(x)~=~\left\{
\begin{array}{cc}
eV , & x<-d/2+eV/F \\
F(x+d/2), &  -d/2+eV/F <x<d/2 \\
U=Fd, & x>d/2%
\end{array}%
\right.   \label{Potential}
\end{equation}
We obtain for quasi-particles flowing perpendicular to the barrier
($p_{y}=0$):
\begin{equation}
P(\epsilon _{0})~\simeq ~\exp \left\{ i\frac{2}{\hbar }\left[ \int_{\frac{%
\epsilon _{0}-\Delta _{R}}{F}}^{\frac{\epsilon _{0}}{F}}p_{+}dx+\int_{\frac{%
\epsilon _{0}}{F}}^{\frac{\epsilon _{0}+\Delta _{R}}{F}}p_{-}dx\right]
\right\} ~,  \label{Newint}
\end{equation}%
where the complex momenta $p_{\pm }$ are determined by the
condition of the quasi-energy conservation
\begin{equation}
\epsilon _{0}=Fx+\tilde{\epsilon}_{e}(p_{\pm }) \label{Newint-mom}
\end{equation}%
Calculating the integrals in Eq. (\ref{Newint}) we write the
transmission probability of the quasiparticles $P(\epsilon _{0})$
as (for a particular case as $\epsilon_0 \geq (U-\epsilon_0)$)
\begin{equation}
P(\epsilon _{0})\simeq  \exp \left[ -\frac{\pi \Delta
_{R}^{2}}{\hbar vF}\right] , (U-\epsilon _{0}) >\hbar \omega /2~,
\label{Tunnsupp}
\end{equation}
%
where the gap $2\Delta_R$ should be taken from Eq. (\ref{a4}).  The Eq. (\ref%
{Tunnsupp}) shows that the external radiation of the frequency
$\omega <2(U-\epsilon _{0})/\hbar $ strongly suppresses the
transmission of the quasiparticles flowing perpendicular to the
junction.

In the opposite case of a large frequency $\omega $ or small energy $%
\epsilon _{0}$, $\omega >2\epsilon _{0}/\hbar $, the resonance
condition, Eq. (\ref{a100}), cannot be fulfilled, the spectrum
remains gapless and the quasiparticle transmission is not
suppressed. In this case the transition occurs from the electronic
quasi-band with $n=1$ to the hole quasi-band with $n=0$ (see the
gray line in Fig. 3a).

There is also a peculiar regime as $(U-\epsilon_0) <\hbar \omega
/2 < \epsilon _{0}$. The interband tunneling for such
quasiparticles is \emph{forbidden }, $P~\simeq ~0$ . Indeed, the
quasi-particles (electrons) on the left side of the junction
starting from the conduction quasi-band have to arrive in the
forbidden one on the right side of the junction (see Fig. 3b).


\begin{figure}[tbp]
\includegraphics[width=3.3in,angle=0]{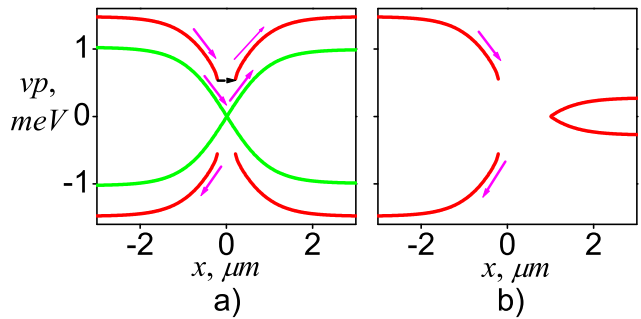} %
\caption{The phase trajectories $p(x)$ of the Hamiltonian
$\hat{H}_{eff}$ corresponding various cases of interband tunneling
a)tunneling through the dynamic gap $\Delta_R$ (thick solid line,
the values of parameters were $U_0=2 meV$, $\epsilon_0=1 meV$, and
$\hbar \omega=0.5 meV$ ); the perfect transmission through the
point $p=0$ (gray solid line, $U_0=2 meV$, $\epsilon_0=1 meV$ and
$\hbar \omega=1.8 meV$); b)the zero transmission case (
$U_0=1.2 meV$, $\epsilon_0=1 meV$ and $\hbar \omega=0.5 meV$). The
$d=1\mu m $ and $\Delta_R=0.2 meV $ were also chosen.}
\label{PhaseTr}
\end{figure}

In experiments, the suppression of the quasiparticle transmission
manifests itself as a strong EF induced increase of the resistance
of the $n$-$p$ junction at small transport voltages $V<[\epsilon
_{0}-\hbar \omega /2]/e$.  The full stationary CVC of the $n$-$p$
junction in the presence of the EF is determined by the elastic
channel, i.e the energies of quasi-particles on the left and right
sides of the junction are equal. Therefore, we can write the
standard expression for the current $I$ flowing through the
ballistic $n$-$p$ junction as \cite{Tunn}
\begin{equation}
I=\frac{4eL}{(2\pi \hbar )^{2}}\int d\epsilon dp_{y}P(\epsilon ,p_{y})\left[
\tanh \frac{\epsilon -eV}{k_{B}T}-\tanh \frac{\epsilon }{k_{B}T}\right] ~~.
\label{Tunncurrent}
\end{equation}%
Here, $\epsilon $ is the quasi-energy of the electrons and $P(\epsilon
,p_{y})$ is the quasi-energy $\epsilon $ and $p_{y}$ dependent transmission
of the quasiparticles through the junction. The coefficient $4$ in Eq. (\ref%
{Tunncurrent}) is due to the spin and valleys degeneracy of the
quasi-particle spectrum in graphene. In the case of a wide junction, when
the width of the graphene sample $L$ satisfies the inequality $L~\gg ~\sqrt{%
\hbar vd/\epsilon _{0}}$, the transmission $P(\epsilon ,p_{y})$ can be
written as
\begin{equation}
P(\epsilon ,p_{y})~\simeq ~P(\epsilon )\exp [-\pi vp_{y}^{2}/(\hbar F)],
\label{a5}
\end{equation}%
where $P(\epsilon )$ is determined by Eq. (\ref{Tunnsupp}). Calculating the
integral over $p_{y}$, and taking the limit of low temperature $T$, we
reduce Eq. (\ref{Tunncurrent}) for the current $I$ to the form
\begin{equation}
I=I_{0}\int_{\epsilon _{0}}^{\epsilon _{0}+eV}\frac{d\epsilon }{\epsilon _{0}%
}P(\epsilon )~~,  \label{Tunncurrent-2}
\end{equation}%
where $I_{0}=\frac{eL\epsilon _{0}}{(\pi \hbar
)^{2}}\sqrt{\frac{\hbar F}{v}} $ is the characteristic current
flowing through the $n$-$p$ junction in the absence of the EF
\cite{Falko}.

Although the exact shape of the CVC is determined by diverse
factors, e.g., by the pre-exponent in Eq. (\ref{Tunnsupp}), and
therefore, by the particular form of the electrostatic potential,
temperature etc., we argue
that the EF with the frequency $\omega <2\epsilon _{0}/\hbar $ leads to the $%
N$ type of the CVC (see Fig. 2). Indeed, as the transport voltage
$V$ is less than the characteristic value $V_{0}=[\epsilon
_{0}-\hbar \omega /2]/e$ the quasiparticle current $I$ flows due
to the interband tunnelling with the probability determined by Eq.
(\ref{Tunnsupp}) (see, the region $1$ on the CVC in Fig. 2 and
corresponding schematic of the process in Fig. 3a). However, in
the voltage region $V_{0}<V<\min \{2V_{0},\epsilon _{0}\}$ the
current $I$ starts to decrease  due to the presence of
quasi-particles whose propagation is forbidden (see, the region
$2$ on the CVC in Fig. 2 and the schematic in Fig. 3b). The drop
of the current becomes
especially deep as $2V_0<\epsilon_0$ and the radiation frequency is in the particular range $%
\epsilon _{0}<\hbar \omega <2\epsilon _{0}$. In the voltage region
$V>2V_{0}$ the current increases with the voltage $V$ because
there is a possibility to propagate with the perfect transmission
for quasi-particles possessing a small momentum $p<\hbar \omega
/(2v)$ (see the region $3$ on the CVC in Fig. 2 and the schematic
in Fig. 3b).

Finally, we address the question of experimental conditions necessary to
observe the predicted effects. An $n$-$p$ junction with the typical width $%
d~\simeq ~1\mu m$ in a graphene sample has to be fabricated. An
external radiation containing the component parallel to the
junction of a moderate intensity $S$ has to be applied. We
emphasize that EF need not be linearly polarized. The EF
suppresses the quasi-particles transmission through the $n$-$p$
junction in the range of the frequencies of EF $\omega ~\leq ~\epsilon _{0}$%
. This means that for the Fermi energy $\epsilon _{0}~\simeq ~0.02eV$ (this
value corresponds to doping levels of a graphene monolayer $n\leq
10^{11}cm^{-2}$ \cite{NovGeim,Zhang}), the EF in the far-infrared region
with the frequency less than $10^{13}Hz$ provides a strong decrease of the
quasi-particle transmission with the intensity $S$ of the radiation.
Choosing an even smaller external frequency $~\simeq ~2\cdot 10^{12}Hz$ and
the width of an $n$-$p$ junction $d=1\mu m$ one may use the radiation with a
moderate intensity $S~>0.4~W/cm^{2}$ to observe this effect. Notice here
that the effect is also reduced at small frequencies $\omega <\epsilon _{0}\sqrt{%
v/(\hbar d\epsilon _{0})}$ because in this case the transport through the $n$%
-$p$ junction is determined by electrons with large values of $p_{y}>p_{x}$.

In conclusion, we have demonstrated that radiation of a moderate intensity $%
S $ having the component parallel to the $n$-$p$ junction in
graphene leads to a pronounced suppression of the quasi-particle
transmission through the junction. This effect occurs due to
formation of a non-equilibrium dynamic gap between electron and
hole bands in the quasi-particle spectrum as the resonant
condition $\omega =2v|\mathbf{p}|$ is satisfied. The value of the
gap can be controlled by variation of the intensity $S$ of an
external radiation. Propagation of quasiparticles is possible due
to the non-equilibrium interband tunnelling. This specific type of
the tunnelling is determined by
the initial energy of electrons $\epsilon _{0}$ and the frequency of EF $%
\omega $ and, as a result we obtain an $N$-type of CVC. The
suppression of the quasiparticle transmission may allow to control
confinement of electrons in diverse structures fabricated in
graphene, like, e.g., $n$-$p$-$n$ transistors, single electron
transistors, quantum dots, etc., by variation of the intensity $S$
and frequency $\omega $ of the external radiation. We hope that
the predicted effect will find its application to future
electronic devices based on graphene.

We would like to thank P. Silvestrov and A. Kadigrobov for useful
discussions and acknowledge the financial support by SFB 491 and
SFB Transregio 12.

\end{document}